# The Ancient Astronomy of Easter Island: The Mamari Tablet Tells (Part 3)

Sergei Rjabchikov[1]

[1]The Sergei Rjabchikov Foundation - Research Centre for Studies of Ancient Civilisations and Cultures, Krasnodar, Russia, e-mail: srjabchikov@hotmail.com

## Abstract

One of main events on Rapanui before the cruel war (several wars) between the Hanau Momoko and Hanau Eepe was the annual election of the bird-man. Plainly, the local priest-astronomers, the descendants of immigrants from the Andean region of South America, could predict the day of the vernal equinox not only with the aid of the so-called "sun stones," but also counting the phases of the moon. The elements of the mathematical calculations of the astronomers have been deciphered. The astronomical basis of the Ahu Atanga (A Tanga) has been decoded, too.

## Introduction

The civilisation of Easter Island is famous due to their numerous ceremonial platforms oriented on the sun (Mulloy 1961, 1973, 1975; Liller 1991). One can therefore presume that some folklore sources as well as *rongorongo* inscriptions retained documents of ancient priest-astronomers.

Guy (1990) as well as Kelley and Milone (2011: 339-340, figure 11.1) repeated basic ideas of my decipherment of the Mamari calendar record (Rjabchikov 1989: 123-125, figures 2 and 3).

## The Month of the Vernal Equinox as the General Calendar Event of the Bird-Man Cult

The natives waited for the emergence of sooty terns every year in September (*Hora-nui* chiefly). Servants of warriors specially lived on the islet Motu Nui that time. One of them who first found the bird's egg swam with it to the island. His lord was proclaimed as the new bird-man.

At the sacred village Orongo, where the festival was conducted, several stones, the so-called "sun stones" stood. Indeed they were an archaic calendar device. It served in order to determine precisely the dates of equinoxes and solstices (Ferdon 1961; 1988). Besides, at Mataveri, another centre of the bird-man cult, the stone calendar with incised lines was discovered; one of lines denoted the day before the vernal equinox of 1775 A.D., another line denoted the day of the summer solstice of the same year (Liller 1989; the interpretation in Rjabchikov 2016a: 1, table 1).

The local priest-astronomers determined the month *Maru* (*Maru*), the time of the winter solstice (June), as the first month; thus, the month *Hora-nui* (September chiefly), the time of the vernal equinox, was the fourth one (Rjabchikov 1993a: 133-134).

## The Inception of Our Studies

Consider a segment of the Mamari (C) tablet, see figure 1.

Ca 10: 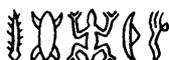

Figure 1.

Ca 10: **25-28 69 3 58** … *Hunga Moko hina (mara, marama) tahi*… The Lizard (the moon/night *Hiro*), the first moon, is hidden…(Rjabchikov 1989: 123-124, figure 2, fragment 3).

Here Old Rapanui *hunga* means 'to hide,' cf. Maori *huna* 'ditto.' Of course, the form *hunga* was quite possible because of the alternation of the sounds *n/ng*. It is an indirect evidence of such a reading. But the direct evidence exists also: the Rapanui expression *hunga raa* signifies 'morning twilight,' cf. Rapanui *raa* 'the sun.' So, the phrase means 'the sun is hidden.'

### The New Mathematical and Astronomical Notes on the Mamari Tablet

As was highlighted earlier, in the calendar record (Ca 4-9) and in the local rock art glyph **68** *hono*, *ono* (to add; to join) played the role of the addition operator (Rjabchikov 2017: 1-4, figures 1 and 2).

Consider now another fragment of the Mamari tablet, see figure 2 (cf. figure 1).

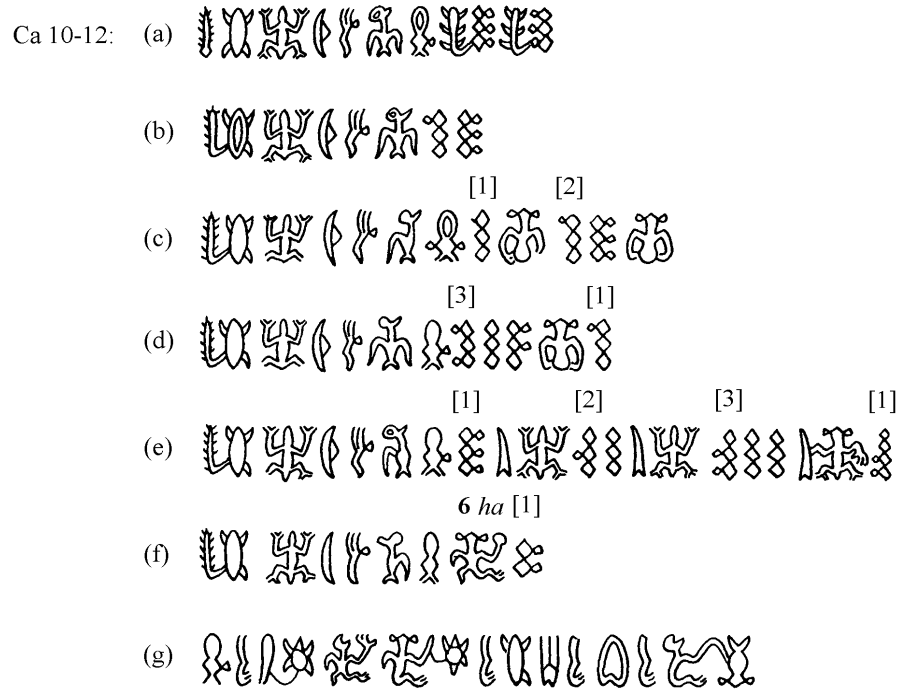

Figure 2.

Ca 10-12: (a) **25-28 69 3 58 44 73 51-17-51-17**
*Hunga Moko marama tahi, taha; he ke tea, ke tea.*
The first moon Lizard (= *Hiro*) is hidden; (it is) the disappearance; the light is hidden, the light is hidden.
(b) **25-28 69 3 58 44 17-17**
*Hunga Moko marama tahi, taha; teatea.*
The first moon Lizard (= *Hiro*) is hidden; (it is) the disappearance; (it is) the bright light.
(c) **25-28 69 3 58 44 73 17 68 17 17 68**
*Hunga Moko marama tahi, taha; he tea, ono, tea, tea, ono.*
The first moon Lizard (= *Hiro*) is hidden; (it is) the disappearance; the first light, add, the second light, add.
(d) **25-28 69 3 58 44 73 17 17 17 68 17**
*Hunga Moko marama tahi, taha; he tea, tea, tea, ono, tea.*
The first moon Lizard (= *Hiro*) is hidden; (it is) the disappearance; the third light, add, (yet one) light.
(e) **25-28 69 3 58 44 73 17 5 69 17 17 5 69 17 17 17 5 69 17**
*Hunga Moko marama tahi, taha; he tea atua Moko, tea, tea atua Moko, tea, tea, tea atua Moko, tea.*
The first moon Lizard (= *Hiro*) is hidden; (it is) the disappearance; the first light of the god Lizard (= *Hiro*), the second light of the god Lizard (= *Hiro*), the third light of the god Lizard (= *Hiro*), (yet one) light.
(f) **25-28 69 3 58 44 73 6 17**
*Hunga Moko marama tahi, taha; (h)e ha tea.*
The first moon Lizard (= *Hiro*) is hidden; (it is) the disappearance; the fourth light.
(g) **73 43 4 7 6 6-7 43 28 1 43 47 43 6 97**

*He ma atua, tuu a Hatu, ma nga Tiki, ma ava, ma amu.*
The god is coming, *Hatu* (= *Tiki-te-Hatu*) is coming, the egg of *Tiki* is coming, the elevation (of the egg) is coming, the food (figuratively) is coming.

Old Rapanui *taha* means 'disappearance,' cf. Rapanui *taha* 'to set (of the sun).' Old Rapanui *ke* means 'to set (of the sun),' cf. Rapanui *keke* 'ditto.' Old Rapanui *tea* signifies 'white; clean; clear; the sun; light; day; shine,' cf. Rapanui *tea* 'white' and *otea* 'dawn' < *oo tea* 'the whiteness enters.'

Old Rapanui *ma* means 'to come; to go' (Rjabchikov 2012a: 565), cf. Maori *ma* 'ditto.' Glyph **7** *tuu* corresponds to Rapanui *tuu* 'to come' (Rjabchikov 1993b: 18). The verbal particle *he* (a rare grammatical article in *rongorongo* inscriptions) precedes one of the verbs *ma* in segment (g). Old Rapanui *nga* 'shell of an egg' corresponds to Maori *nganga* 'shell; husk.'

Let us try to decode the name of the place *Tanema* located on the western coast of Pitcairn Island. It could be the name invented by Tahitian wives of the Bounty's sailors. In this case, *Tanema* reads *Tane ma* (the sun god *Tane* is going). West was not only the direction of the setting sun, but also the certain mark of the distant Society Islands.

Now one can study the mathematical and astronomical aspects of that record. The sentence "The first moon Lizard (= *Hiro*) is hidden" repeated several times denote that the priest-astronomers counted the new moons (months). Segments (a) and (b) were presented to stress that a solar eclipse was possible, but it did not occur. According to segments (c) – (f), the natives waited for the fourth month called *Hora-nui* (September chiefly) when sooty terns arrived on the off-shore islet Motu Nui.

In segment (c) we read: [1] 'the 1st day' '+' [2] 'the 2nd day' '+.' In segment (d) we read: [3] 'the 3rd day' '+' [1] 'the 1 day.' In segment (e) we read: [1] 'the 1st new moon,' [2] 'the 2nd new moon,' [3] 'the 3rd new moon,' [1] '(the next) day (= the next new moon).' In segment (f) we read: *(h)e ha tea* 'the 4th day = the 4th new moon (the new moon of the 4th month *Hora-nui*).'

**The Number Four in the Rapanui Beliefs**

In different Rapanui folklore texts the number four is presented to designate four days, four parts of the island, four ghosts, four men, four winds, four seasons (*tau*), four islets as the symbols of the abundance, power, and great force (Blixen 1974: 3-7; Englert 2002: 30-31, 102-107, 174-175, 192-193, 268-269; Felbermayer 1971: 64-79, 91-93; Métraux 1940: 65-67, 260-261, 378-381, 385-386, 389; Routledge 1998: 279-280). One can also mention the Rapanui expressions *vero, vero, vero, vero* (four spears = a great spear) and *e te manu vae e ha* (the bird with four legs = the warrior with rapid legs) (Rjabchikov 2016b: 5; 2017: 27).

In a manuscript by A. Gaete (the informant Leonardo Pakarati, 1973; see Fedorova 1978: 47-48, 368), where a variant of the local Creation Chant is put down, four small lizards are mentioned. Since inscribed versions of the Creation Chant could be put down in manuals in the *rongorongo* schools, one of exercises could be devoted to four lizards.

I have disclosed the following record on the Great Santiago (H) board as a recommendation for teachers in the royal *rongorongo* school called *Hare Titaha* (see details in Rjabchikov 2012a: 568-569, figure 8), see figure 3.

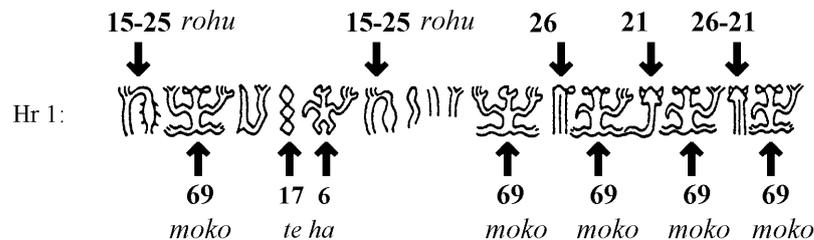

Figure 3.

Hr 1: **15-25 69 5-15 17 6 15-25 52 5-15 69 26 69 21 69 26-21 69**

*Rohu MOKO atua roa: **te ha**! Rohu, hiti: atua roa MOKO, **ma** (or **mo-**) MOKO ᵒ**ko**, MOKO **ma-**(or **mo-**)ᵒ**ko**, MOKO!*
Create (the words) 'LIZARD, the great god' **four (times)**! Create, lift (the words from this line to another): 'The great god LIZARD, **ma-** (or **mo-**) LIZARD ᵒ**ko**, LIZARD **ma-** (or **mo-**) ᵒ**ko**, LIZARD!'

    Students wrote four lizard signs **69** *moko* together with the quasi-syllabic signs **26** *mo*, *ma*, *maa* and **21** *oko*, *ko* on lessons in the royal *rongorongo* school. The number of the lizard glyphs was take down as the conjunction of glyphs **17 6** *te ha* meaning 'four', perhaps 'fourth' as well.

### The Mamari Tablet as the Main Lesson Book in the Royal *Rongorongo* School

The analysis of different fragments from the known inscribed boards shows that the Mamari board was the source of many segments there. Two of such examples are presented in Rjabchikov 2012a: 567-568, figure 6; 568-569, figure 7. Even the fragment in figure 3 can be a variant of segments (c) – (f) of figure 2. Three additional parallels are gathered in appendix 1. Per my studies, the tablets Tahua (A) and Keiti (E) as well as perhaps the Small Santiago (G) and London (K) tablets were in the royal school during the reign of king *Kai Makoi* the First. The Aruku-Kurenga (B) tablet, "Tablette échancrée" (D), Small Vienna (N), Small Washington (R), Great Washington (S), Great Santiago, Great St. Petersburg (P) and Small St. Petersburg (Q) tablets were inscribed in that school during the reign of his son, king *Nga Ara*.

### How King *Nga Ara* Checked *Rongorongo* Experts

In accordance with Routledge (1914-1915), during the annual readings of *rongorongo* records at the royal residence of Anakena king *Nga Ara* sat in front of the house called *Hare Papa Marama*, and experts *tangata rongorongo* stood. The ceremonial platform was called *Ahu Vai Mamari Mor[e]*. The monarch "taught *rongorong[o] taki manu tau*."
    Having compared the drawing of Anakena area made by Routledge and the modern maps, it is apparent that *Hare Pa-pa Marama* (The House where (there are) records – the school) on that drawing is the so-called house of king *Hotu-Matua*. The abbreviation EIL means 'Easter Island line.' The platform has the name *Ahu Naunau* now.
    I conclude that the king read the Mamari tablet during such meetings. The house Hare Pa-pa Marama = Hare Titaha (the House at the former border between the Hanau Eepe and Hanau Momoko). The name *Vai Mamari More* signifies '(The place where) the records (*more*) of (the tablet) Mamari are given (*vaai*).' The phrase *taki manu tau* means 'the appearance of birds – the season,' cf. Rapanui *taku* 'to predict,' *tataku* 'to add; to count,' Maori *taki* 'to lead along; to recite.' So, all the segments of figure 2 inscribed on the Mamari board could be the crucial text recited many times during such annual readings.

### The Mamari Board in the *Rongorongo* School of the Tupa-Hotu Tribe

After the tablet Mamari (*kohau o te ranga*) was stolen from king *Nga Ara*, a Tupa-Hotu man used it as a primer (Routledge 1998: 249). Routledge (1914-1915) recorded these words from the recitations of that tablet: *Karoa veke tit huna vere awe mahia hia lange ure*.
    I have reconstructed the text as follows: *Ka roa, veke, ti(a)-ti(a):* **huna vere** *(h)au. E mahi(a)-hi(a) ranga ure*. 'Add, preserve, write (the words) many times: **huna vere** from the tablet. Write (the text of the tablet) *ranga* giving the abundance.'
    Old Rapanui *roa* means 'to increase; to grow,' cf. Rapanui *roroa* 'to grow tall.' Old Rapanui *veke* means 'to preserve,' cf. Mangarevan *veke* 'ditto.' Old Rapanui *kohau* (*ko hau*) and *hau* mean 'tablet,' cf. Rapanui *hau* 'string;' maybe the term was related to the *quipu*, the knot script of the Old Peruvians (see Rjabchikov 2012a: 566, figure 2, fragment 3; 569, figure 9). Old Rapanui *hi* means 'to carve; to cut,' cf. Rapanui *hiahia* 'to saw' (Rjabchikov 2012b: 15, figure 5).
    Old Rapanui *vere* signifies 'rain,' cf. Rapanui *vere hiva* 'drizzle' (cf. also the name of the Rapanui ghost of the rain *Mata Varavara*; the alternations of the sounds *e/a* were quite possible). Hence, the expression *huna vere* correlates with glyphs **25-28 69** *hunga* (= *huna*) *Moko* from the Mamari tablet.

King *Nga Ara* used this record on the board during his school lessons as well as during annual *rongorongo* readings, it was well known, and therefore the teacher from the Tupa-Hotu tribe used the same record in the beginning of his lessons.

**The Expression *Huna Moko* was Inscribed on Tablets in the *Nga Ara*'s *Rongorongo* School**

Consider the parallel records on the Great St. Petersburg and Aruku-Kurenga tablets (cf. the former interpretations in Rjabchikov 2012b: 21, figure 13; 22, figure 14), see figure 4.

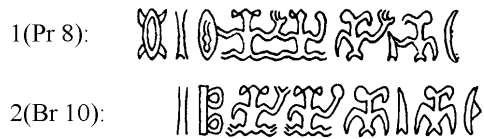

Figure 4.

1 (Pr 8): **28 4 25/30 69-69 44 5-44 3**
*Nga ti(a): huna Mokomoko. Taha. Titaha marama.*
(There are) many records (*nga tia*): (the expression) *huna Mokomoko* '[the moon] of the Lizard (the first moon *Hiro*) is hidden.' Turn (the tablet). (It is the lesson in) the school of Hare Titaha.
2 (Br 10): **4-4 39-39 69-69 44 5-44 3**
*Ti(a)-ti(a) raaraa Mokomoko. Taha. Titaha marama.*
(There are) many records (*tia-tia*): (the expression) *raaraa Mokomoko* 'the days of the Lizard (the first moon *Hiro*) are hidden.' Turn (the tablet). (It is the lesson in) the school of Hare Titaha.

In the first record the particle *nga* (Plural) was written down (glyph **28**) before the word *ti(a)*. In the second case the doubling of the same word *ti(a)* was used to put the plural form down.
Certainly, glyphs **25 30** (or **28**) **69** (or **69-69**) read *huna* (= *hunga*) *Moko* (= *Mokomoko*, *vere*, *vara*, *verevere*, *varavara* etc.).

**The Cross-Readings of Some Glyphs**

1. Consider the brief record on the London *rei-miro* (J), a possible royal ornament, see figure 5.

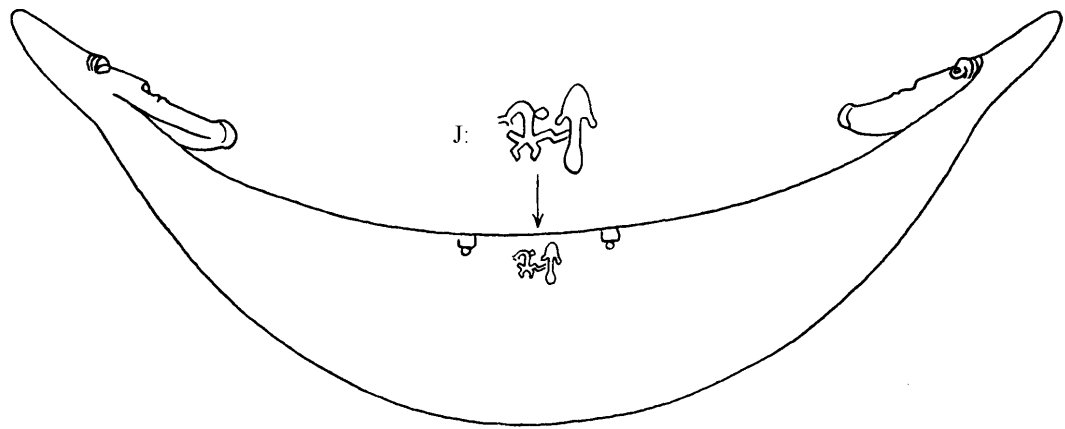

Figure 5.

J: **49a 22** *(Ariki) mau ao*. The king has authority.

2. Consider the record on the Santiago staff (I), see figure 6.

I 1: 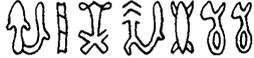

Figure 6.

**I 1: 22 (102) 4 49c 59-33 (102) 28 6-39-6-39** *Ao atua, (ariki) mau, kaua Nga Araara.*
*Nga Araara* (= *Nga Ara*) who is the lord, the king (and) the progenitor has authority (Rjabchikov 2012a: 566-567, figure 3).

Notice that the name of king *Nga Ara* was also retained as *Nga Araara erua* (Routledge 1914-1915). The latter expression means 'The old man (*e rua*) *Nga Araara* (= *Nga Ara*)' (Rjabchikov 2009: part 5). It is clear that the staff once belonged to that ruler.

Consider now a motif on the skull of a man aged 45 to 49 years from Easter Island (Owsley et al. 2016: 263, figure 14.5). It is glyph **22** *rapa*, *ao*. In this case, the sign reads *ao*. Rapanui *ao* means 'ruler, chief, king' (Englert 1948: 378-385) and 'victory' (Métraux 1940: 378-381) = 'authority' indeed. Thus, it is the cranium of a chief or king.

3. Consider the records on the Santiago staff, Honolulu (T) and Keiti tablets, see figure 7.

1(I 11): 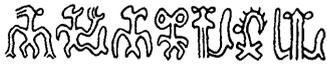

2(T 4): 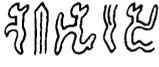

3(I 14):

4(Ev 7): 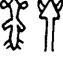

Figure 7.

1 (I 11): **6 6-(102)-44 60 26 (102) 69var 49 4-26 (102) …** *Ha hata mata Momoko (ariki) mau tuma* (= *tumu*)… The tribe of the Hanau Momoko, the king of (that) family (*ariki mau tuma*) appeared…
2 (T 4) **19-26 72 (102) 58 6 …** *Kumaa MANU taiho…* BIRDS *kumara, taiho…*
3 (I 14): **6-21 133** *haoko koreha* eels *koreha haoko* [cf. Rapanui *koreha* 'eel; worm' and Quechua *kuru* 'worm']
4 (Ev 7): **131 (= 132) 21** *koreha oko* eels *koreha oko* (= *haoko*)

The first text describes the arrival of the Hanau Momoko, the forefathers of the Miru tribe, on Easter Island. The second text reports the names of birds *kumara* 'Oestrelata incerta or leucoptera' and *taiho* 'petrel' presented in lists of the local bird species (Métraux 1940: 18; Barthel 1978: 149). Eels of six-foot long were called *koreha haoko* (Brown 1996: 186). It is known that the natives sometimes caught only eels for their ruler *Kauaha* (Englert 1948: 378-385). Hence, a number of eels were a good gift for king *Nga Ara*, according to the third text. Rapanui *oko* means 'ripe,' and therefore the name *haoko* (*ha oko*) means 'to become ripe.' In the fourth text, the name of the eel *haoko* is written as *oko*. This segment is a part of two sentences where several eel species are listed (Rjabchikov 2011a: 11, figure 13).

4. Consider the parallel records on the Great Santiago and Aruku-Kurenga tablets, see figure 8.

1(Hv 10): 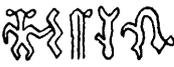

2(Bv 9): 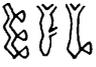

Figure 8.

1 (Hv 10): **6 52 26 9var 13var (= 132) …** *Ha hiti, ma niu Tuna…* 'The nut grew out, came up from the Eel *Tuna*…'
2 (Bv 9): **52 9 13var (= 132) …** *Hiti niu Tuna…* 'The nut grew out from the Eel *Tuna*…'

Glyph **9** *niu* represents the sprouted nut. In different contexts this sign can denote the nut and the palm. Both texts are local variants of a Polynesian myth about the coconut palm which grew out from the head of the Eel *Tuna* (Métraux 1940: 323). Old Rapanui *tuna* 'eel' correlates with Maori *tuna* 'ditto'.

5. Consider two records on the Great Santiago tablet, see figure 9.

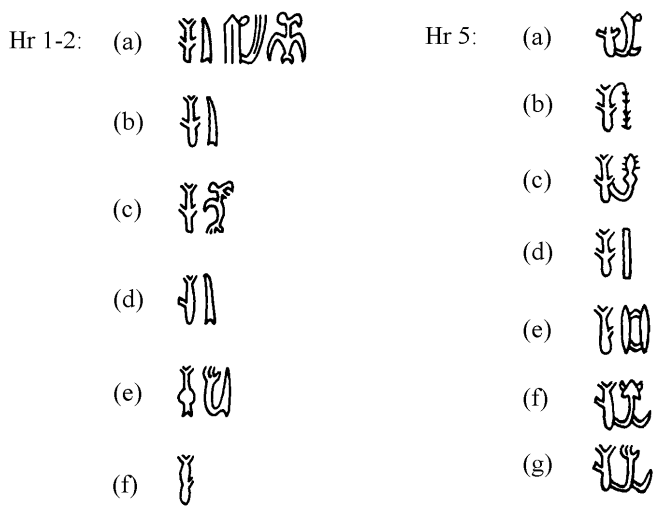

Figure 9.

Hr 1-2: (a) **9 5-26 32 44** (b) **9 5** (c) **9 19** (d) **9 5** (e) **9var 15 5** (f) **9 …**
(a) *Niu tuma (= tumu) uta*, (b) *niu hati*, (c) *niu ki*, (d) *niu hati*, (e) *niu roa, hati*, (f) *niu…*
(a) (There were) nuts in the upper part of the palms, (b) nuts were cut down, (c) (there were) many nuts, (d) nuts were cut down, (e) big nuts were cut down, (f) (there were) nuts…

Glyph **9var** in segment (e) includes the small circle (the image of the nut as the determinative). Old Rapanui *ki* means 'many; great,' cf. Hawaiian *ki* 'supreme, great.'

Hr 5: **…** (a) **9 13** (b) **9 24** (c) **9 25** (d) **9 4** (e) **9 30 149-149 30** (f) **9 21 5** (g) **9 15 5**
… (a) *niu Tuna*, (b) *niu ari*, (c) *niu hua*, (d) *niu hati*, (e) *niu ana (hotuhotu) ana*, (f) *niu oko hati*, (g) *niu roa hati*.
… (a) nuts from the Eel *Tuna*, nuts, nuts-fruits, nuts which were cut down, numerous nuts, ripe nuts which were cut down, big nuts which were cut down.

If the final signs in segments (f) and (g) are **3 (= 61)**, they read *marama*, *mara* and even *rama*, cf. Rapanui *rama* 'nut.'

6. Consider the records on the Aruku-Kurenga and Honolulu tablets, see figure 10.

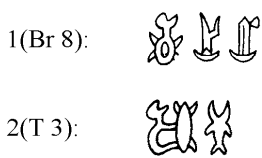

Figure 10.

1 (Br 8): **6-28 3 5-5 3 26** *Hanga hina (= marama) titi, hina (= marama) maa.* 'The full moon, the bright moon is moving.'
2 (T 3): **6-28 11** *Hanga Pakia* 'The bay of Seals.'

In the first text glyphs **6-28** *hanga* 'to move' correspond to Maori *anga* 'ditto.' In the second text the same glyphs read *hanga* 'bay.' Earlier I reconstructed PPN **paki* 'whale; seal' on the basis of Rapanui *pakia* 'seal,' Maori *pakakā* 'whale; seal' (< **paka*) and even Ainu *pakuy* 'nerpa' (Rjabchikov 2014a: 163). Now one can add Pileni *pakeo* 'shark' to this list: PPN **pak(i,e,a)* 'whale; seal; shark.'

7. Consider the parallel records on the Berlin and Aruku-Kurenga tablets, see figure 11.

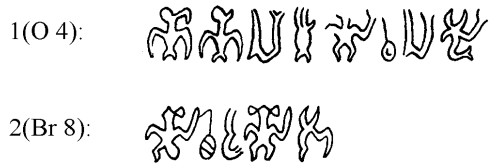

Figure 11.

1 (O 4): **44-44 5-15 12 15 6-103 5-15 11 …** *Tahataha atua roa Ika roa hope, atua roa Pakia (Mango)…* 'The great god of the large fishes as the food, the great god of sharks (seals, dolphins, whales etc.) were hidden.
2 (Br 8): **6-103 43-6 11var** *hope maha (= mango) MANGO (PAKIA, NIUHI etc.)* 'sharks (seals etc.) as the food.'

In the first text the large sea creatures (sharks, tuna fish and so forth) prohibited for the catching during winter and spring months (till the month *Tangaroa-uri*, October chiefly) are described. In both texts Old Rapanui *hope* 'food' is presented.

8. Consider the parallel records on the Santiago staff and Tahua tablet, see figure 12.

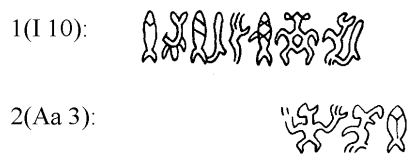

Figure 12.

1 (I 10): **12 (123) 12 (102) 15 12 6-19 (102)** *ika, ika roa, ika aku* 'fishes, large fishes, the fish *aku*.'
2 (Aa 3): **6-19 12** *aku IKA* 'the fish *aku*.'

Conceivably, the fishing ground Te Aku Renga situated not far from the royal residence Anakena (see Barthel 1978: 194) is described in both texts.

9. Consider the record on the Keiti tablet, see figure 13.

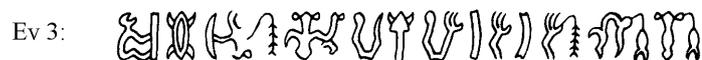

Figure 13.

Ev 3: **6-4 28 3 45 24 68 56 50var 50-21 50var 15-4-15-4 15-24 44b 12 56 12** *A atua Nga Hina (= Vina) Pu ai: hono po; i hiko, i roturotu ro ai tua IKA, poo IKA.* (It was) the goddess '*Vinapu* (the place);' (it was) adding nights [at this solar and lunar observatory]; (the fishermen) snatched (and) collected fishes from the deep sea, the fish *poo-poo*.'

The words *nga hina* mean 'the many moons' literally = 'the great (almost full or full) moon.' Old Rapanui *hiko* means 'to snatch,' cf. Maori *hiko* 'ditto' (cf. Rjabchikov 2011a: 11, figure 13).

This inscription tells of the ceremonial complex including the platforms Ahu Vinapu 1 (known as Tahiri 'The Lift' also) and Ahu Vinapu 2. The place name *Vinapu* reads *Vina Pu = Hina Pu* 'The Bearing moon goddess *Hina*' (Rjabchikov 1990: 22-23). This conclusion was based on the interpretation of the female personage *Vivina* as the moon goddess *Hina* (Barthel 1957: 64).

Interestingly, glyph **108b** *(h)iri* (to lift) is inscribed together with lunar glyphs and certain names of nights of moon age on one stone cylinder from that religious complex (Rjabchikov 2001: 219).

At Vinapu fishhooks were recovered in a large number; for Vinapu 1 the dates range between 1516 A.D. ± 100 years; the platform of Vinapu 2 could be made with Old Peruvian technologies (Mulloy 1961). Women from Vinapu married men of the Tupa-Hotu, and some of those men carved statues at the quarry Rano Raraku (Barthel 1978: 280).

Consider a complicated motif (a penis and a vulva as a fertility design) on a skull found at Ahu Vinapu 2 (Owsley et al. 2016: 267, figure 14.10). I suppose that such symbols could have the Old Andean origin (cf. Scher 2010: 280-282, 439, figure 6.31; 442, figure 6.37; the interpretation in Rjabchikov 2017: 17). It was the designation of the supreme Andean (Old Peruvian) god *Tiqsi* (the Rapanui sun god *Tiki*).

The text about the Vinapu statues put down on the Mamari tablet will be examined in one of next parts of this work.

10. Consider the record on the Santiago staff, see figure 14.

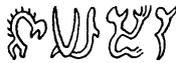

Figure 14.

1 (I 14): **62var 61 (102) 6 105** *Too Hina a moa*. '(The moon goddess) Hina took fowls.'

Apparently, they were offerings for the moon goddess. As a parallel, examine these data. Statue UU-15 lies not far from one of platforms at the ceremonial area Urauranga te Mahina; the remains of human beings and chickens were excavated there (Ayres et al. 2014). I suggest that it was the monument representing the dark moon (Rjabchikov 2016a: 7-8). The people and fowls were, without any doubts, sacrifices for that moon goddess.

11. Consider the parallel records on the Tahua and Small Washington tablets, see figure 15.

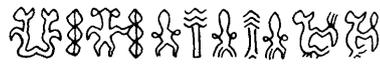
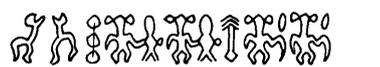

Figure 15.

1 (Aa 6): **19-19 17 6 17 73 6/33 73 6/33 73 6 1? 6 30** … *Kiki: te ho, te e, atu, e, atu, e, a Tiki* (?), *a ana*… 'Say: (the sign) *ho*, (the sign) *e*, learn: (the sign) *e*, learn (the sign) *e*, (the god) *Tiki* (= the sign of abundance), many times…
2 (Ra 2): **19-19 17 6-73-6-73 4/33 6-30-6-30** … *Kiki: te hoehoe, atu, a ana, a ana…* 'Say: (the signs = the word) *hoehoe* (paddle), learn (it) many times, many times…

In both cases the instructions of one lesson in the *rongorongo* school were put down: students wrote glyphs *ho* and *e* many times in order to learn the word *hoe* (paddle). Notice that the grammatical article *te* (glyph **17**) could be missed in some cases.

12. Consider the record on the Keiti tablet, see figure 16.

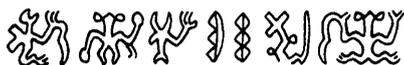

Figure 16.

Ev 8: **72 50 6-4 11 18-17 70 4-50 69** *Manu hi hotu poki Teatea pu tuhi Moko* 'The Bird of the sunbeams (= the white booby *kena*) is a creature-child (*hotu poki*) produced (*pu*) by the Whiteness = the bright sun (*Teatea*) (and) by the *Tu(h)i Moko* = the Absence-Lizard (the darkness, disappearance, night, eclipse, rain, winter).'

The parallel text is presented in the Creation Chant (Métraux 1940: 320-322): *Vie Moko ki ai ki roto kia Tea ka pu te kena* 'The woman *Moko* [Lizard] by copulating with Whiteness produced the booby.'

The old man *Ure Vae Iko* recited the Creation Chant by heart (Thomson 1891: 514ff), it is therefore apparent that the text had been once inscribed on a tablet in the royal *rongorongo* school. Different sentences of that chant with some variations could be inscribed on other boards in that school.

13. Consider the record on the Mamari tablet, see figure 17.

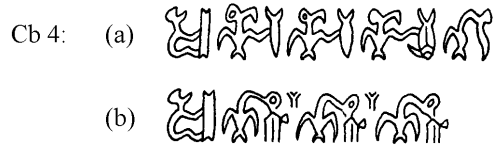

Figure 17.

Cb 4: (a) **6-4 44-16 44-16 44-16 44b** *A atua Ta(h)a Kahi, ta(h)a Kahi, ta(h)a Kahi tua,* (It was) the god 'The shore of the Tuna Fish, the shore of the Tuna Fish, the shore of the Tuna Fish of the open sea,'
(b) **6-4 44-26-15 44-26-15 44-26** *a atua Tamaroa, Tamaroa, Tama.* (it was) the god '*Tamaroa = Tangaroa, Tamaroa = Tangaroa, Tama(roa) = Tanga(roa).*'

It is the description of the ceremonial platform Ahu Atanga (= A Tanga) in the western part of the island (Rjabchikov 2014a: 173; 2014b: 4). The name *A Tanga* means '(The deity) *Tanga* (*Taanga, Tama, Tangaroa, Tamaroa*). It is well known that this platform is located at the almost most northern rim of Easter Island. Moreover, the *ahu* points to north very precisely (Liller 1991: 274, table 1). Notice that the sun moved (and moves) from east via north to west each day. Hence, the *ahu* was a solar observatory.

Since the god *Tangaroa* was the sun deity in the Society Islands and in the Western Polynesia, the remnants of that cult could be retained in the Rapanui astronomical notations. Thus, the platform oriented on the midday sun was named after the sun deity. It is common knowledge that the Miru tribe and the neighbours believed in the supreme deity *Tangaroa*.

14. Consider the record on the "Tablette échancrée," see figure 18 (the drawing is corrected).

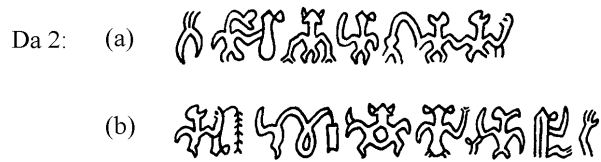

Figure 18.

Da 2: (a) **2** (= the inverted glyph) **62-56 6-21 6 54 6-6 15** *Ina topa, hakaha Kai-haha roa.* The great *Kai-haha* did not appear (and) did not breathe.
(b) **62-56 24 91 4 6-25 68 5 44-26 6det.var 15** *Topa ari(k)i Taoraha atua, hakahono atua tama* HUMAN *roa.* The chief (called) *Taoraha*-the lord appeared, (he) was the next lord; he was a son (of *Kai-haha*). (Cf. Rjabchikov 1993a: 139-140, appendix 2, figure 5, fragment 74; the transliteration, reading and translation have been corrected.)

It is the copy of a text taken from a tablet that once belonged to the Hanau Eepe (the Tupa-Hotu tribe). The name of the chief *Kai-haha* (glyphs **54 6-6**) means 'The mouth eats.' *Taoraha* 'The whale' (glyph **91**) was the next ruler, a son (*tama roa* meaning 'great child' here) of *Kai-haha*. His name as *Taoraha-Kaihahanga*, i.e. *Taoraha*, a son of *Kai-hahanga* (= *Kai-haha*), is registered in the lists of the rulers of Easter Island (Métraux 1940: between 90-91, table 2).

Glyphs **49** *(ariki) mau* 'king' and **22** *ao* 'authority; ruler; king' are not found in the decoded record. Hence, both rulers commanded certain districts of the island in fact.

In my opinion, *Kai-haha* (*Kai-hahanga*) was called *Kaina = Kainga* (The eater), he was a chief of the Hotu-Iti tribal union (the Hanau Eepe) and lived near Tongariki (Tonga Ariki); *Taoraha* was known as *Huriavai* [*Huri a vai*] 'A son from the water' (Thomson 1891: 529-531; Routledge 1998: 282-288).

Glyph **6det.var** represents an arm and a leg of a human being. This sign was used as a determinative (human being; different human deeds). Here glyphs **6-21** are the causative prefix *haka-*.

15. Consider the record on the Aruku-Kurenga board, see figure 19.

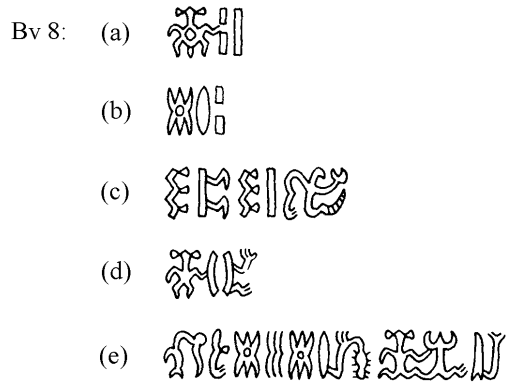

Figure 19.

Bv 8: (a) **6 29 4** *Ha rutu*: (It was) recited:
(b) **7 30 29** *Tuu Ana, Rua.* The Shine (and) the Sunset came (*tuu*).
(c) **52 4 29 52 4 31** *Hiti atua Rua, hiti atua Make*. The god of the Sunset appeared, the sun god *Makemake* appeared.
(d) **6 57 6det.var** *A tara* HUMAN BEHAVIOUR: (It was) the incantation:
(e) **19 43 7 33 7 30 15-25 19 23 5-**15 *Ku ma, tuu tai, tuu ana rohu, Moko huru atua roa*. The ocean (water etc.) came (*ma, tuu*), the appearing shine (and) the great god 'The Lizard in the feathers' came (*tuu*).'

In Manuscript E (Barthel 1978: 311) such an incantation is preserved: *Ko Ene, ko tuu tai, ko taka hiti, ko Ruhi*. I have translated it as follows: '(It was) the Shine (*Ene = Ana*), the ocean came (*tuu*), the sun (*taka*) appeared, (it was the chthonian god) *Ruhi*.'

The god *Ruhi* of the sunset and the new moon was the incarnation of the first statue of the ceremonial platform Ahu Tongariki in the Rapanui myth "*Ko Ruhi*" (Heyerdahl and Ferdon 1965: figure 164; the interpretation in Rjabchikov 2010a).

In both versions of the charm the dual model of the world was reflected: two main antagonists were the sunshine and the darkness in different conditions. The ocean (water) was a substance between the day and night in the local beliefs.

Old Rapanui *tara* means 'prayer, incantation, recitation,' cf. Tahitian *tara* 'to be saying a prayer.' Glyph **58** *tai*, *tahi* represents the billow, cf. the shape of glyph **33** *vai*, *ua*. It is clear that in some cases glyph **33** reads *ngaru*, *tai* etc. The spirit *Moko huru* (*Ruhi, Rua, Hiro, Hiva kara rere* 'The Blackness – the wings are flying') was the symbol of the chthonic component of the sun god *Maui-tikitiki* (the sun during the new moon when a solar eclipse could fall out).

16. Consider the record on the "Tablette échancrée," see figure 20 (the drawing is corrected).

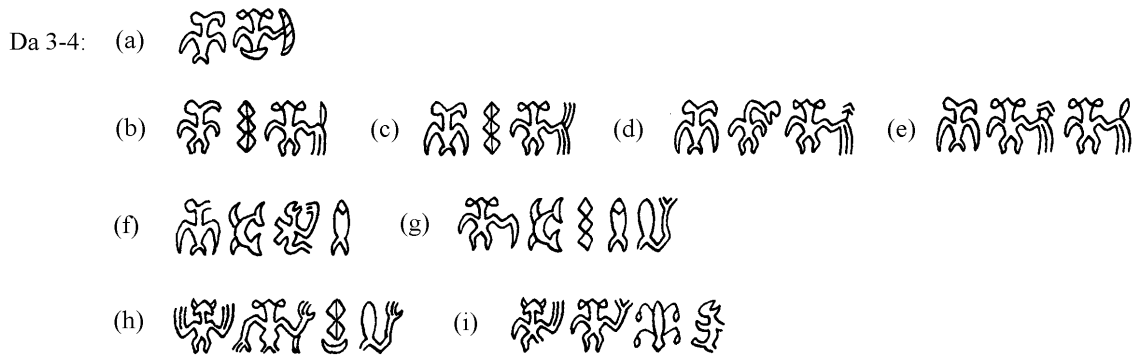

Figure 20.

Da 3-4: (a) **44 3 6 3** *Taha marama ha marama:* Four (*ha*) months passed [after the beginning of the year] (= it happened in the fourth month Hora-nui, September chiefly):
(b) **6 44 17 6-32 3** (or **30**) *Ha taha te Haua Hina.* The moon *Haua* (= the 13rd moon *Atua*) passed.
(c) **44 17 6-32 32** *Taha te Hauaua.* The moon *Haua* (= the 13rd moon *Atua*) passed.
(d) **44 62 6-32 33** *Taha, to Hauaua.* The moon *Haua* (= the 13rd moon *Atua*) passed.
(e) **44 6-32 4/33 6-32 30** *Taha Haua Atua, Haua ana*. The moon *Haua* (= the 13rd moon *Atua*), the abundant *Haua* (or the brilliant *Haua*) passed.
(f) **44 8 72 12** *Taha* (or *Tava*) *Matua manu Ika.* (King) *Hotu*-Matua (glyphs **44 8**) was the bird (the warrior metaphorically) of the Fish (= the god *Tangaroa* or *Tinirau*).
(g) **6 8 17 12 48-15** *A Matua te Ika Uri..* (King) *Matua* (arrived) in the month *Ika Uri* (= *Tangaroa Uri*, the fifth month, October chiefly).
(h) **32 6-21 32 68 15 3 17 48-15** *Ua Aka ua honga roa marama* (or *mara*) *tea, uri.* A great priest (*tahonga*) of the bright (and) dark moon (once) was at the dwelling (*ua*) Aka (= Akahanga).
(i) **6-21 32 6-25 19** *Aka ua ahu Kioe.* The dwelling (*ua*) Aka (= Akahanga) was the platform (*ahu*) of the Rat (= king *Hotu-Matua*).

Here some notes of mine on this inscription have been collected.

(a) The fourth month *Hora-nui* was designated with the word *ha* 'fourth.' Besides, the second glyph **3** *marama* (month) is decorated with four lines, maybe it was the determinative (the number four).

(b)-(e) The months were counted in this record from the almost full moon known as *Atua* 'The goddess' and *Haua* (the moon goddess *Hina*'s name) to the same lunar phase.

In segments (b) and (c) the grammatical article *te* (glyph **17**) is presented, and in segments (d) and (e) it is omitted. I demonstrated before that homogeneous parts of a sentence – nouns – are preceded by glyphs **17** *te* = the definite articles *te* (Rjabchikov 1987: 361, 364-365, figure 2, fragment 3; 1993a: 128, figure 2, fragment 9; 128-129, figure 3, fragments 8 – 11; 130, figure 4, fragment 5; 2016b: 1, figure 2, segment (c); this paper, p. 9, figure 15, fragment 1). Furthermore, with the help of the methods of structural linguistics the sentences "V – glyph **17** *te*, i.e. the article *te* – S" were found, where the V was one verb existing in various forms (glyphs **6-4** *hotu*, **6-6-4** *hohotu*, **6-4-6-4** *hotuhotu* 'to bear a fruit'), and the different S were the terms related to plants (Rjabchikov 1993a: 129, figure 3, fragments 1 – 7).

Old Rapanui *to* means 'to set (of the sun or the moon),' cf. Tuamotuan *tō* 'ditto' and Rapanui *tonga* (< *to-nga*) 'winter, season of rains.'

(a)-(g) The text tells of king *Hotu-Matua* who began his voyage on the fourth month *Hora-nui* (September chiefly) and reached the Easter Island on the next month *Tangaroa-uri* (October chiefly).

(h) When the Hanau Eepe controlled the whole island, the residence of the great priest-astronomer was situated at Akahanga. Because of his status his name is recorded in the lists of Rapanui kings: *Tuhunga roa* 'The great priest' (see Métraux 1940: between 90-91, table 2). In Manuscript E (Barthel 1978: 85) this passage is put down: *Akahanga a hare hakamahangahanga*. I have translated it as follows: '(It was) Akahanga, (it was) the house where (somebody) taught knowledge (*haka-ma-hanga-hanga*).'

The place name *Akahanga* (cf. glyphs **6-21** *Aka*, and glyph **32** *UA* as a determinative) contains the word *aka* and the suffix *-hanga*, cf. Rapanui *aka* 'root' and Samoan *a'a* 'fibres of a root; family connections' (see also Rjabchikov 2009: part 2). Moreover, the name of the place Akapu, where every new king lived (Métraux 1940: 132), contains the same word *aka*. Thus, the expression *Aka pu* means 'The root bears (descendants, offsprings),' cf. Old Rapanui *pu* 'to produce; to bear.'

(i) A rat was the incarnation of the soul of king *Hotu-Matua* (Englert 1948: 74). Per the local non-realistic traditions, the monarch was buried at (near) the Ahu Akahanga (Thomson 1891: 510).

## Conclusions

One of main events on Rapanui before the cruel war (several wars) between the Hanau Momoko and Hanau Eepe was the annual election of the bird-man. Plainly, local priest-astronomers, the descendants of immigrants from the Andean region of South America, could predict the day of the vernal equinox not only with the aid of the so-called "sun stones," but also counting the phases of the moon. The elements of the mathematical calculations of the astronomers have been deciphered. The astronomical basis of the ceremonial platform Ahu Atanga (A Tanga) has been decoded, too.

# Appendix 1

### Example 1

Consider such parallel records, see figure 21.

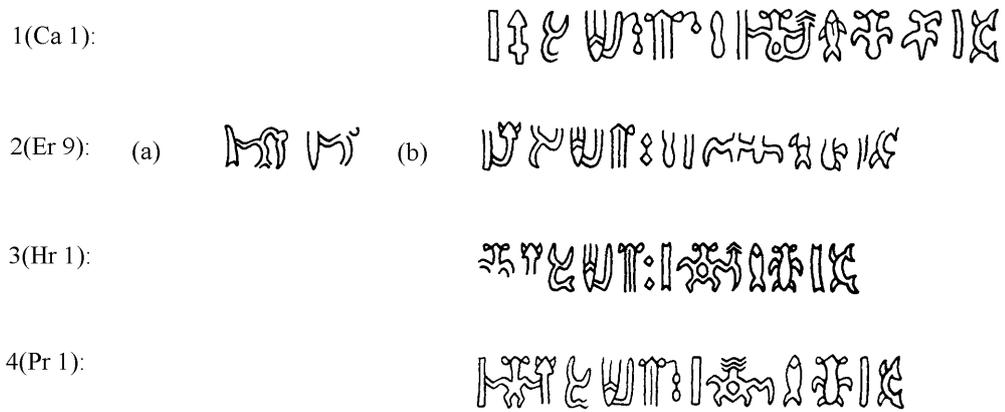

Figure 21.

Crucial words from these records are collected in figure 22.

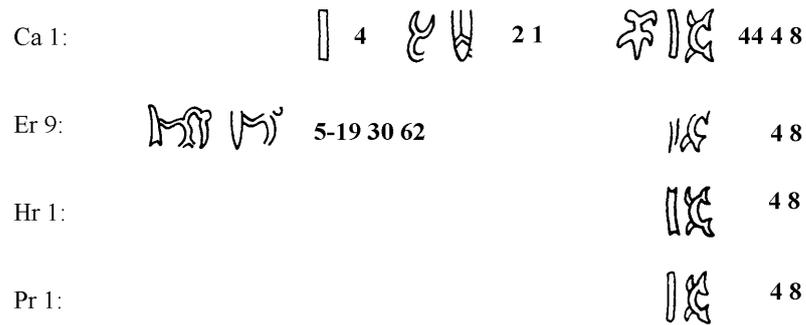

Figure 22.

**4** *atua* 'god; goddess; gods'
**2** *Hina* 'the moon goddess'
**1** *Tiki* 'the sun god'
**44** *Taha* 'the Frigate Bird (*Tiki*, *Makemake*; *Tangaroa*);' *manu* 'bird' or *manu-tara* 'sooty tern' (figuratively) in some cases
**4 8** *Atua-Matua* (or *Atua-Metua*) 'the god *Tane* (*Tiki*)'

These words are written down in the beginning of a number of sentences on the Keiti tablet:

**5-19 30 62** *tuki (= tuku) ana to* 'strike (= write) many times, add' (Rjabchikov 2013a).

Old Rapanui *tuku* 'to strike = to write,' a variation of Rapanui *tuki* 'to strike,' is mentioned in Rjabchikov 2012a: 565. The alternation of the sounds *i/u* was quite possible. In Manuscript E this text (in Roman letters) was crossed out: *i tuki mai ki roto kia au* (Barthel 1978: 350). The text means: '(it was) struck (= written) on a tablet (*au* = *hau*)' (it is the translation of mine). Thus, at least one unknown *rongo-rongo* board was a source for that manuscript.

# Example 2

Consider such parallel records, see figure 23.

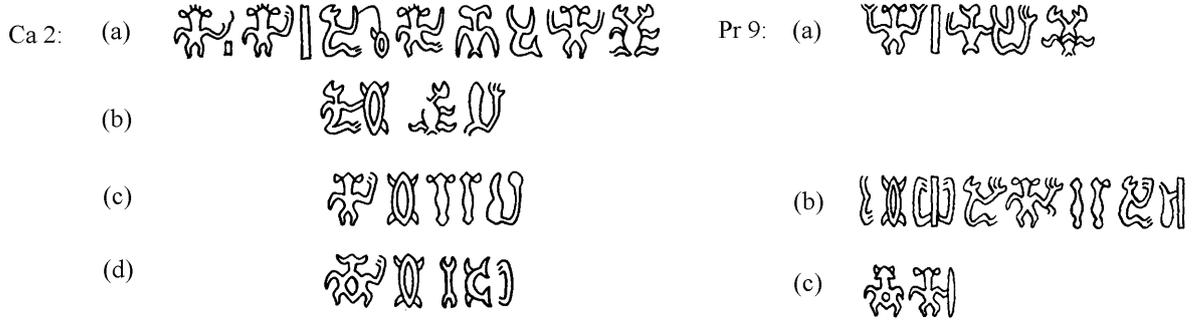

Figure 23.

Crucial words from these records are collected in figure 24.

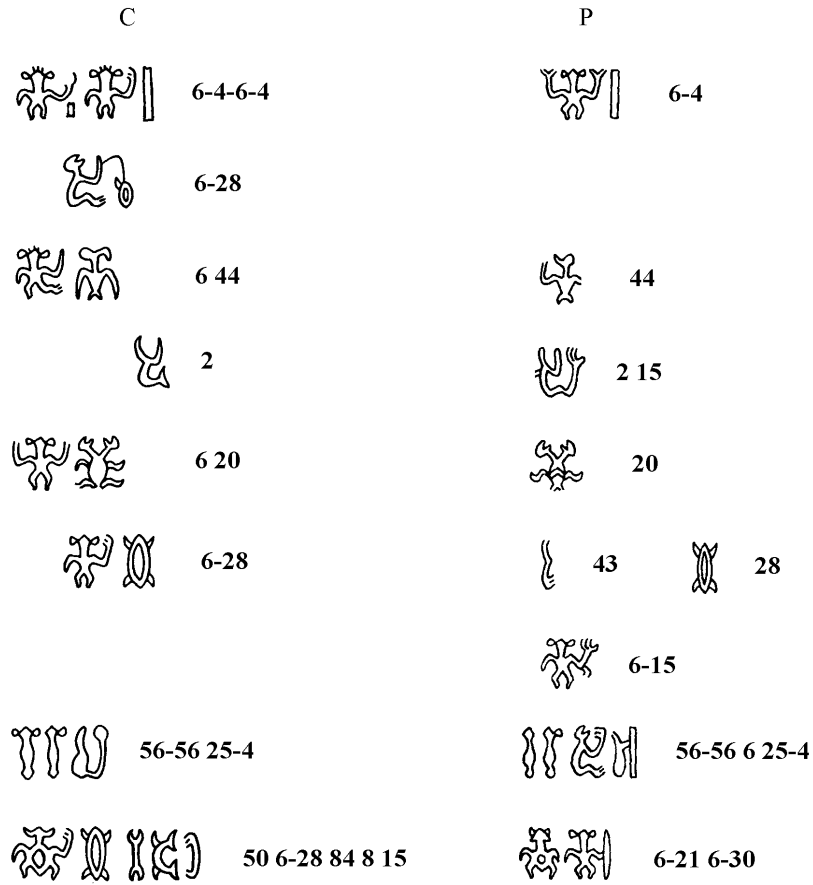

Figure 24.

These words are deciphered as follows.
Glyphs **6-4-6-4** read *hatuhatu* (*hotuhotu*) 'to appear,' glyphs **6-4** read *hatu* (*hotu*) 'ditto.'
Glyphs **6-28** read *hanga* 'to move.'
Glyphs **6 44** read *a Ta(h)a* 'the moon/night *Rongo Tane*,' glyph **44** reads *Ta(h)a* 'ditto.'
Glyph **2** reads *Hina* 'the moon goddess; the moon,' glyphs **2 15** read *Hina roa* 'the great moon goddess; the big or full moon.'
Glyphs **6 20** read *a Pikea* (or *Ungu*) 'the Crab' (the Rapanui-Polynesian moon goddess *Hina*; the Old Peruvian moon goddess *Killa* who acted as a crab), glyph **20** reads *Pikea* (or *Ungu*) 'ditto.'
Glyph **43** reads *ma* 'to go; to come.'
Glyph **28** reads *anga* 'to move.' Glyph **28** reads not only *nga*, but also *anga* (Rjabchikov 2016c: 3).
Glyphs **6-15** read *Hora* 'the season of the months *Hora-iti* and *Hora-nui*, August-September chiefly.'
Glyphs **56-56 25-4** read *Popo hutu* 'The Surf, the Wave' (= *Paryaqaqa*, the name of the Old Peruvian storm god), glyphs **56-56 6 25-4** read *Popo a hutu* 'ditto.'
Glyphs **50 6-28 84 8 15** read *hi, hanga ivi Matua roa* 'the ancestor 'The great Father' shone, moved' (here the name of the Old Peruvian god-creator *Tiqsi*, the Polynesian god-creator *Tiki* was encoded).
Glyphs **6-21 6-30** read *hakahana* '(it was) the heating.'

In two cases mentioned above glyph **6** *ha, a* reads *a* (the grammatical article of personal names). As was substantiated earlier (Rjabchikov 1988), this glyph as the particle *a* introduced the personal names in the genealogy on the Small Santiago tablet. The genealogy had been discovered by Butinov and Knorozov (1957: 15: table 7, fragments 1 – 6), see figure 25, fragment 1. In the parallel segment on the Small St. Petersburg tablet the article *ko* of personal names (glyph **21var**) is presented, see the same figure, fragment 2.

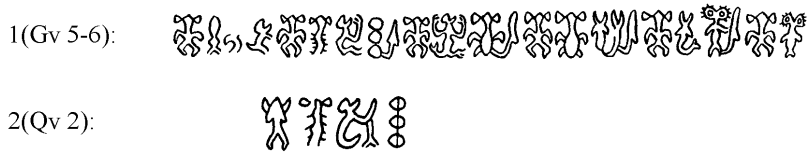

Figure 25.

On the New York wooden figurine of a bird-man (text X) glyphs **6-46 6-7** *hana Hatu* 'the heat (the sun) of the god *Tiki-te-Hatu*' are inscribed together with glyph **1var** *Tiki* (the sun god) and glyph **28** *nga* 'egg' (Rjabchikov 2010b: 9-10, figure 2, fragment 1), see figure 26. The observations of the motion of the sun along the ecliptic in the spring-time (during the growing warmth of the sun) were the astronomical basis of the bird-man cult. No doubt the ancient priest-astronomers watched the sun constantly.

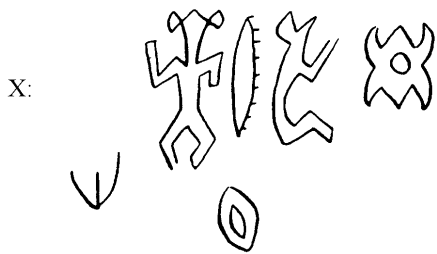

Figure 26.

# Example 3

Consider such parallel records, see figure 27.

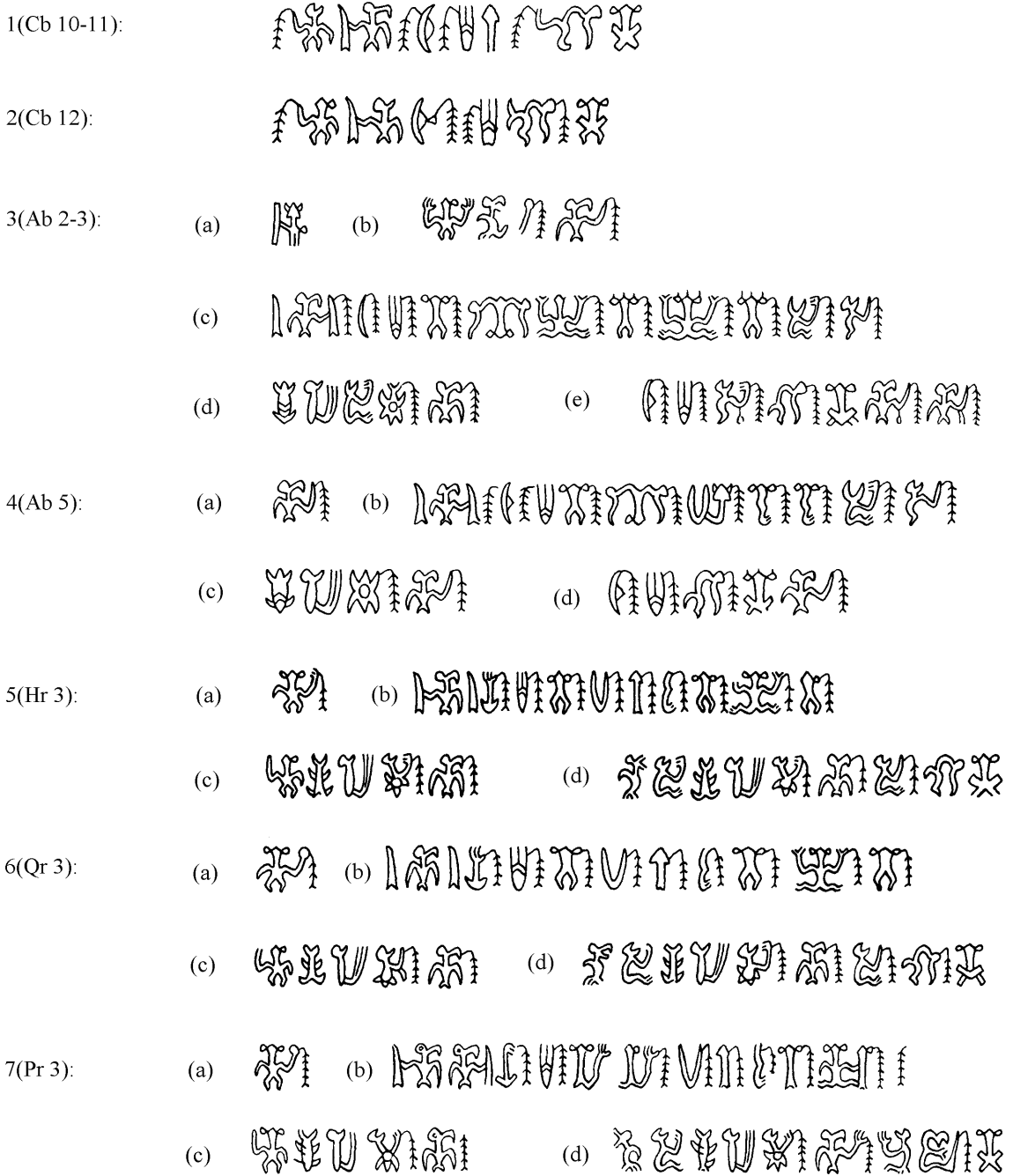

Figure 27.

Crucial words from these records are collected in figure 28.

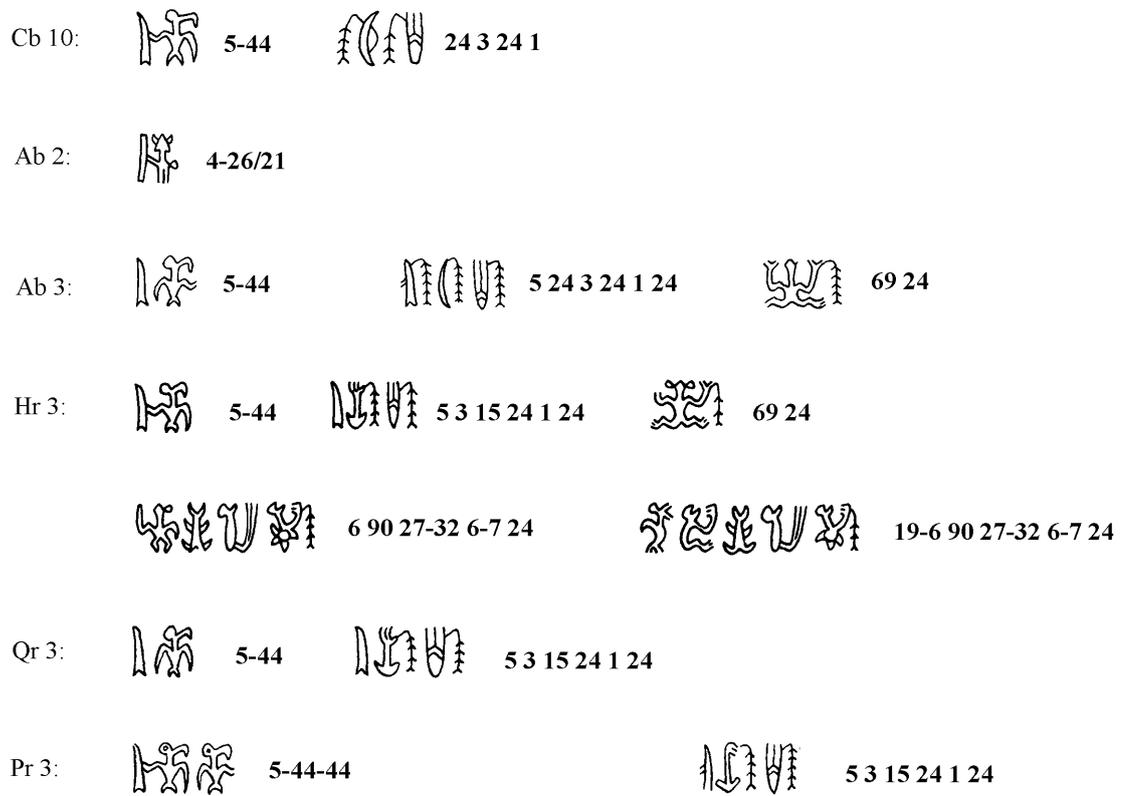

Figure 28.

Glyphs **5-15** *Titaha* (The Border) denote the landmark (the big house) at the territory of Anakena. The other segments describe the statues of the ceremonial platform Ahu Naunau. Glyphs **24 3 24 1** read *ai Hina, ai Tiki* 'the place (statue *mo-ai*) of *Hina*, the place (statue *mo-ai*) of *Tiki*.'

Glyphs **4-26/21** *timo ako* 'a pupil strikes = writes' introduce different didactic records on the Tahua, Great Santiago and Great St. Peterburg tablets.

The parallel segments read **5 24 3 24 1 24** *atua-ai: Hina ai, Tiki ai* 'the gods of the places (*mo-ai*): the place (statue *mo-ai*) of *Hina*, the place (statue *mo-ai*) of *Tiki*,' and **5 3 15 24 1 24** *atua: Hina roa ai, Tiki ai* 'the gods: the big *Hina* as a place (statue *mo-ai*), *Tiki* as a place (statue *mo-ai*).' Glyphs **69 24** *Moko ai* 'The Lizard (the chthonic deity *Hiro* known as *Hiva Kara Rere*) as a place (statue *mo-ai*)' denote another statue on that platform. This segment turns out to correspond to the drawing of a large lizard engraved on a stone of this a*hu*.

Glyphs **6 90 27-32 6-7 24** read *Ha: Para raua Hatu ai*. Glyphs **19 6 90 27-32 6-7 24** read *Kuaha: Para raua Hatu ai*. The name of the ghost *Kuaha* [*Ku haha*] means '(Somebody) has seen' is associated with the view and light, cf. Rapanui *ha* 'to gaze' (Rjabchikov 2011b: 6). In the first segment the simplified name of that spirit is written as *Ha* (The View). It was the big-eyed god *Makememe* (*Tiki*, *Maui-Tikitiki*, *Tane*, *Tangaroa*) indeed. The phrase *Para raua Hatu ai* means '(The deity) *Para* together with (the deity) *Hatu* as a place (statue *mo-ai*).' It is the description of statues of the gods *Para* and *Tiki-te-Hatu* stood near each another on the Ahu Naunau. They could be images of the Old Peruvian (Andean) gods *Parya Qaqa* and *Tiqsi*. Notice that the eastern Rapanui tribe Hiti-Uira (the Hanau Eepe, Hotu-Iti) was called *Ure-o-Hei*, too. The first ethnicon means 'The lightning appears,' and the second one 'The Kin of the Storm' (cf. Rjabchikov 2017: 25). Thus, the tribe was named after the archaic storm deity.

Glyphs **5-44-44** *Titahataha* (The Border) are the variant of glyphs **5-44** *Titaha*.

The Old Peruvian (Andean)-Polynesian trans-Pacific contacts could be reflected in some Polynesian string figures also (cf. Rjabchikov 2000). It is well known that the Hawaiian string figure "Winking Eye" (Dickey 1928: 141-142, figure 102) exactly corresponds to the Peruvian string figure "The Chicken Bum" (Elffers and Schuyt 1978: 34). I have distinguished the sign "rhombus" in the string patterns in both cases. In some other Hawaiian string games this symbol was comparable with the Rapanui glyph **67** *pi* (cf. Rjabchikov 1997: 44, 47-48, table). In the Old Peruvian linear script (Rjabchikov 1994: 54, table 8; 2017: 17) the sign "rhombus" reads as Quechua *punchai* 'day' and *inti* 'the sun.'

Glyph **67** *pi* denoting the fertility and abundance is engraved on a Rapanui royal wooden staff *ua* that is housed in the National Maritime Museum, Greenwich, London (Rjabchikov 2013b: 8, figure 2). By the way, glyph **44** *Ta(h)a* (*Tavake, Kena, Manu-Tara, Manu* and so forth) 'The Frigate Bird; The Tropic Bird; Booby; Sooty Tern , Bird; etc.' – maybe the name of king *Hotu-Matua* (*Tava*, *Tavake*, *Nga Tavake*) or another name (e.g., *Ura ki Kena* or *Uru Kena*, or *Te Urua ki Kena* = *Hotu-Matua*?), or an epithet of the god *Tiki* (*Tane* etc.), or a sign of the bird (bird-man) cult, or an ethnicon (cf. Mangarevan *Nga Tavake*, the name of a tribe) – is seen on another Rapanui royal wooden staff *ua* which is housed in the National Museum of New Zealand, Wellington. Such wooden batons with Janus-like heads represent the all-Polynesian god *Tinirau* (Rjabchikov 2014a: 167).

The Old Peruvian sign "rhombus" *punchai* could be a source of the Rapanui glyph **67** *pi* (the gradation of the sounds *u/i* was quite possible). It is still a hypothesis.

## Acknowledgements

I wish to thank Fr. Paul Lejeune for his kind permission to study four excellent *rongorongo* tablets (Mamari, Tahua, Aruku-Kurenga, and Tablette échancrée) in the General Archives of the Congregation of the Sacred Hearts of Jesus and Mary (Rome) in May 2015. I am grateful to Ms. Luana Tarsi for her assistance during that research.

# Appendix 2

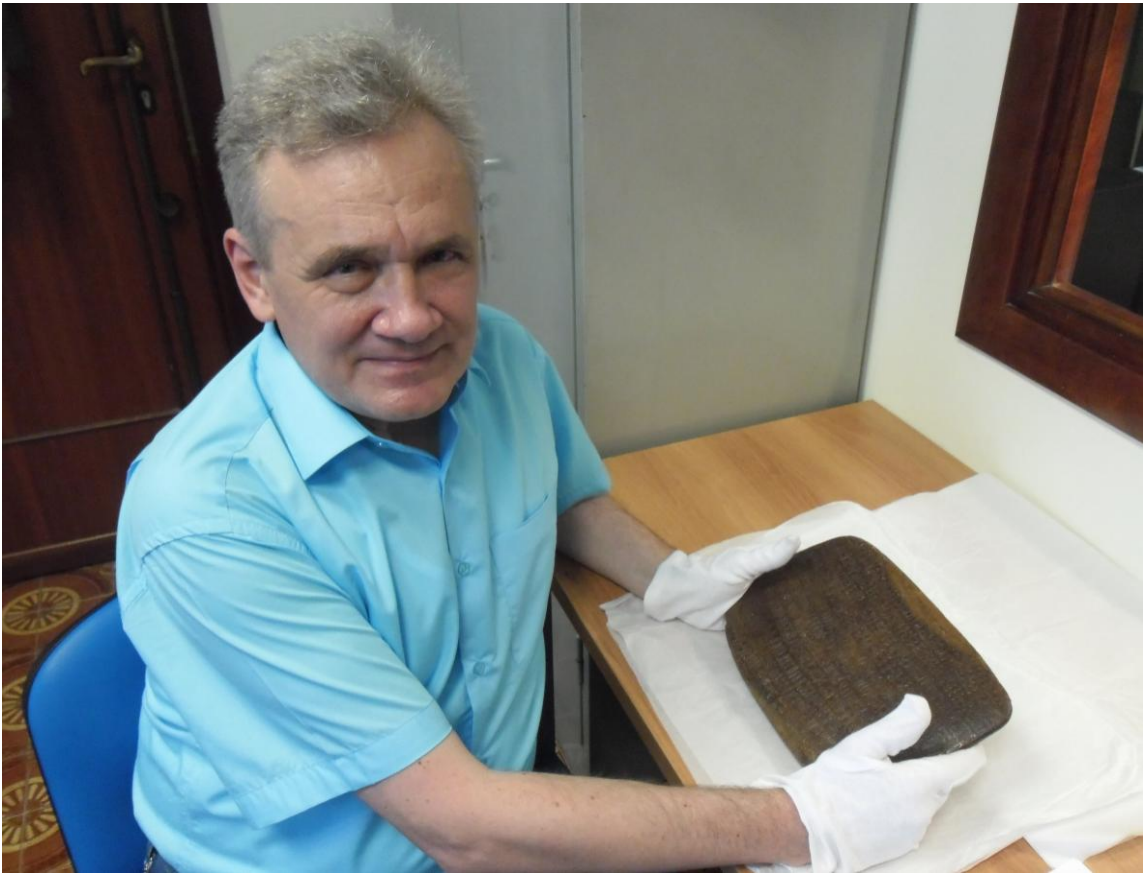

I visited the General Archives of the Congregation of
the Sacred Hearts of Jesus and Mary in 2015.
I was tracing the *rongorongo* glyphs of the Mamari tablet.

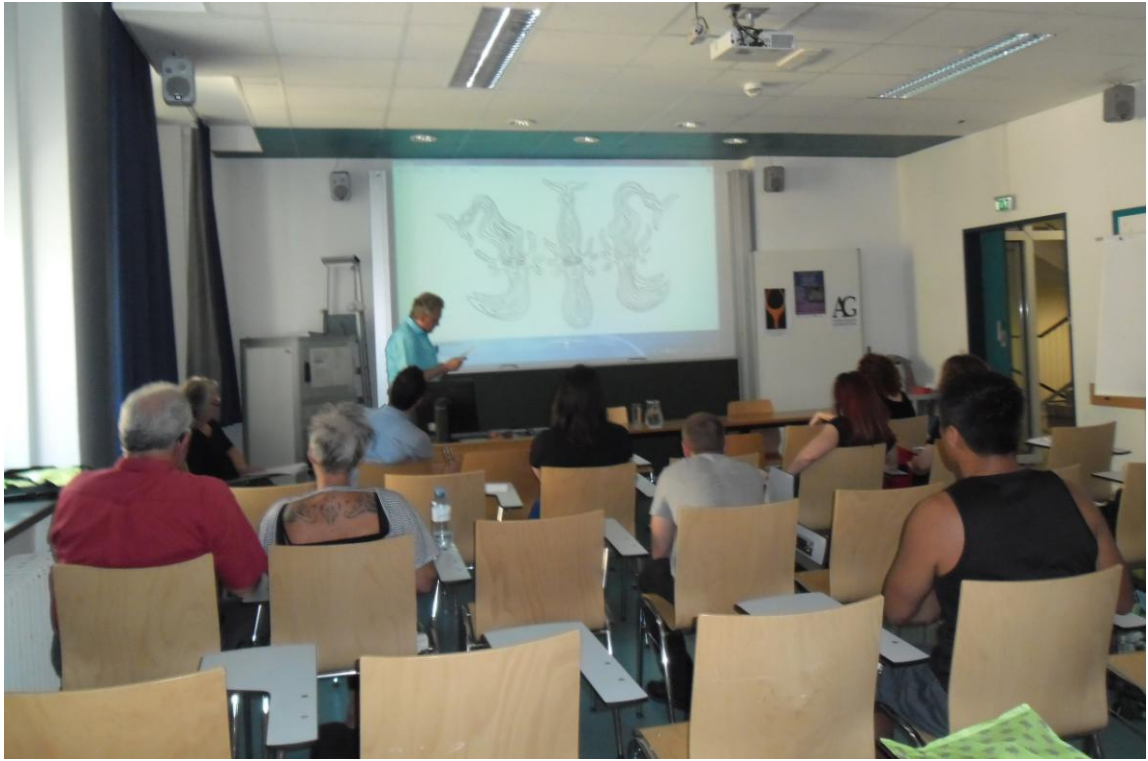

I delivered the lecture "Tama-nui-te-ra, Tangaroa, Tane, Whiro: Remarks on the Maori Pantheon" on a session of the 21st annual conference of the New Zealand Studies Association, University of Vienna, July 3, 2015.

I continued to investigate the phenomenon of the trinity of the archaic Polynesian god *Tama* (Father; Ancestor) known as *Tamaroa* and *Tangaroa*. I displayed on the screen a pattern of three divine whales which embellished the head of a wooden figurine *moai tangata* from Easter Island.

The article written on the base of that report was published:

Rjabchikov, S.V., 2016. Tama-nui-te-ra, Tangaroa, Tane, Whiro: Remarks on the Maori Pantheon. *Polynesia Newsletter*, 6, pp. 2-4.

Additional data about the god *Tama* (T*angaroa*) were published here:

Rjabchikov, S.V., 2016. The *Rongorongo* Script Has Been Deciphered. *Polynesia Newsletter*, 6, p. 5.